\documentclass[12pt,a4paper]{article}
\usepackage{amsmath}
\usepackage{amsfonts}
\usepackage{amssymb}
\usepackage{makeidx}
\usepackage[applemac]{inputenc}
\usepackage{tikz} 

\begin{document}
\sloppy

\newtheorem{definition}{Definizione}
\newtheorem{comment}{Osservazione}
\newtheorem{statement}{Principio}
\newtheorem{theorem}{Teorema}

\title{\textbf{Nonholonomic constrains: why does not the least action principle leads to equations describing the motion consistent with the physical behavior?}}
\author{Umberto Lucia\\ I.T.I.S. `A. Volta', Spalto Marengo 42, 15121 Alessandria, Italy}
\date{}
\maketitle

\begin{abstract}
The least action principle seems not to lead to equations describing the motion consistent with the physical behavior for nonholonomic constrains. Here an answer to this question in proposed.
\end{abstract}

\section{Introduction}
The mechanical systems are running movements which are restricted by constraints, induced by material achievements geometrically expressible as lines, curves, planes, surfaces [\ref{morro}]. In order to characterize the constraint, then, it must be known both its mathematical description (equations that express the constraint) and its physical nature (the forces that express the binding reaction) [\ref{morro}].

In order to determine the spatial position of a system of  $N$ material points, it is necessary to identify the values of $N$ position vectors $\mathbf{r}$, i.e. $3N$ coordinates $r_i, i\in[1,3 ]$ if the system is free or $n\leq 3N$ coordinates if this is the number of system's freedom degrees [\ref{landau}]. If the points representing the system are all possible, then the system is free, while if they are subjected to the restrictions of the constraints they are said to be bound [\ref{morro}], then it may be more convenient not to choose a Cartesian reference system, but may seem more convenient to introduce a system of $n$ generalized coordinates $q$ appropriate to the problem considered [\ref{landau}]. The knowledge of the only generalized coordinates is not sufficient to determine the mechanical condition of a system at a given time, but the values of the generalized velocities is required, i.e. it needs to know the couples $(\mathbf {q},\dot{\mathbf{q}})$, where $\mathbf{q} = (q_1,\dots,q_n) $ and $\dot{\mathbf{q}} = (\dot{q} _1,\dots,\dot {q} _n)$, at the same time [\ref{landau}].

\begin{definition} - 
A constraint is said:
\begin{enumerate}
\item holonomic, any restriction on the possible configurations \emph{[\ref {morro}]} of the system:
	\begin{equation}
	f(\mathbf{q},t)=0
	\label{olonomi}
	\end{equation}
	and it is an integrable relation;
\item nonholonomic, any restriction on the movements possible \emph{[\ref{morro}]} of the system:
	\begin{equation}
	g(\mathbf{q};\dot{\mathbf{q}},t)=0
	\label{anolonomi}
	\end{equation}
	and it is not an integrable relation.
\end{enumerate}
If the nonholonomic constraints represents a holonomic constraint, then it is integrable.
\end{definition}

\section{Holonomic and Lagrangian systems}
\begin{definition} 
- \emph{[\ref{morro}]}
Let $Q$ be a set of points. It is said map of size $n$ on $Q$ an application injective $\varphi: U \subseteq Q \rightarrow \mathbb{R}^n $ with image the open set $\varphi(U)$ in  $\mathbb{R}^n$. The $n$ functions $Q_ {i}:U \rightarrow \mathbb{R}, i\ in [1,n] $ such that $\forall x \in U: \varphi(x) = \big(Q_ {1}(x),\dots,Q_{n}(x)\big)$ are the coordinates associated with the fold $ \varphi$. The $\mathbf{q} = \{Q_ {i}\}_{i \in [1,n]}$ form a local coordinate system on all $Q$. It denotes the fold with the pair $(U,\varphi)$ or $(U,\mathbf{q}) $.
\end{definition}
\begin{comment}
- \emph{[\ref{morro}]}
The introduction of the concept of map allows us to represent the domain of a real open set, so it can be used as a mathematical instrument of analysis.
\end{comment}

\begin{definition} 
- \emph{[\ref{morro}]}
Two maps of dimension $n$, $\varphi_{1}: U_{1}\rightarrow\mathbb{R}^{n}$ and $\varphi_{2}: U_{2}\rightarrow\mathbb{R}^{n}$ are said $C^{k}-$compatible if $U_{1}\cap U_{2}=\emptyset$ or if, when $U_{1}\cap U_{2}\neq\emptyset$, the two following conditions occur:
\begin{enumerate}
\item the sets $O_{1}=\varphi_{1}(U_{1}\cap U_{2})$ and $O_{2}=\varphi_{2}(U_{1}\cap U_{2})$, imagin of the intersection of the two domain on the two maps, are open;
\item the transition function $\varphi_{12}: O_{1}\rightarrow O_{2}$ and $\varphi_{21}: O_{2}\rightarrow O_{1}$, defined as $\varphi_{12}=\varphi_{2}\circ \varphi_{1}^{-1}$ and $\varphi_{21}=\varphi_{1}\circ \varphi_{2}^{-1}$, with $\varphi_{1}$ and $\varphi_{2}$ of class $C^k$ restricted to the intersection $U_{1}\cap U_{2}$.
\end{enumerate}
\end{definition}

\begin{comment}
- \emph{[\ref{morro}]}
The transition functions are application between the two open systems $\mathbb{R}^n$ represented by functions as  $q_{1i}=\varphi_{12i}(q_{1h})$ and $q_{2i}=\varphi_{21i}(q_{2h})$, which allow us to describe a change in coordinates between one map to another one.
\end{comment}

\begin{definition} 
- \emph{[\ref{morro}]}
On the set $Q$ a set of compatible maps is defined as $\mathcal{A}=\{\varphi_{\alpha}: U_{\alpha}\rightarrow \mathcal{R}^{n}; \alpha\in\mathcal{I}\}$, with $\mathcal{I}$ set of indeces with domains $U_{\alpha}$ which are an overlap of $Q$. A set $Q$ with atlas is said differential variety of dimension $n$. If the atlas has all the possible maps, then it is said full or filled or maximum. A differential variety is a set with maximum atlas.
\end{definition}

\begin{comment}
- \emph{[\ref{morro}]}
An atlas allows a topology, so a differential variety is also a topological space.
\end{comment}

\begin{definition} 
- \emph{[\ref{morro}]}
A set of points $\{P_{\nu}, \nu\in\mathcal{B}\}$ is said holonomic if its space of configurations $Q$ has the structure of differentiable variety. Then, $Q$  is saied variety of configurations. The dimension $N$ of $Q$ is said nomber of freedom degree of the system. The coordinats $q_{i}$ related to every maps of $Q$ are said lagrangian coordinates.
\end{definition}

\begin{comment}
- \emph{[\ref{morro}]}
$\forall \nu\in\mathcal{B}, \exists \mathbf{r}_{\nu}: Q\rightarrow E_{3}$, i.e. there exists an application which assigns the position vector $\mathbf{r}_{\nu}$ of the point $P_{\nu}$ to ay configuration od teh system: known the coordinates $q_i$ on $Q$ the applications $\mathbf{r}_{\nu}$ are vectorial functions $\mathbf{r}_{\nu}(q_i)$. Consequently, the velocity is $\dot{\mathbf{r}}_{\nu}=\sum_{i}\dfrac{\partial\mathbf{r}_{\nu}}{\partial q_i}\dot{q}_i$.
\end{comment}

\begin{definition}
- \emph{[\ref{morro}]}
The motion act of a holonomic system is a set of vectors ${(\mathbf{r}_{\nu},\dot{\mathbf{r}}_{\nu}), \nu\in\mathcal{B}}$ such that:
\begin{equation}
\left\{\begin{array}{ll}
\mathbf{r}_{\nu}=\mathbf{r}_{\nu}(q_i)
\\
\dot{\mathbf{r}}_{\nu}=\sum_{i}\dfrac{\partial\mathbf{r}_{\nu}}{\partial q_i}\dot{q}_i
\end{array}
\right.
\end{equation}
\end{definition}

\begin{comment}
- \emph{[\ref{morro}]}
If $Q$ is the configuration variety then the set of the action acts is the tangent variety $TQ$; indeed, the $\dot{q}_i$ of the motion acts are the components of a vector tangent to $Q$ on the coordinates $q_i$.
\end{comment}

\begin{definition}
- \emph{[\ref{morro}]}
A holonomic system is a system of points whose possible configurations in all times are a differentiable variety $\bar{Q}$ of dimension $n+1$, said space-time of the configurations, such that:
\begin{enumerate}
\item there exists a differentiable function $t:\bar{Q}\rightarrow\mathbb{R}$ which assign to any configuration its time;
\item this application is such that  $\forall t\in\mathbb{R}$ the set $Q_t$ of all the possible configurations at the time $t$ is a sub-variaty of dimension $n$;
\item there exists a differentiable variety $Q$ od fdimension $n$ and a diffeomorfism $\varphi : \mathbb{R}\times Q\rightarrow\bar{Q}$ such that in any $i\in\mathbb{R}$ it generates a diffeoemorfism  $\varphi_{t}:Q\rightarrow Q_{t}:q\mapsto\varphi (t,q)$ between the variety $Q$ and the variety $Q_t$. The integer $n$ is the number of degree of freedom and the variety $Q$ is the reference configuration variety.
\end{enumerate}
\end{definition}

A holonomic system is made of constrained or free points. In dynamics, the action of the constrain is a force of reaction, the constrain reaction, on qhich constitutive conditions must be imposed. The smooth constrain is represented by the orthogonality between the constrain reaction and the constrain itself. For a holonomic system, for a forces configuration and system $\mathbf{F}_{\nu}$ applied to every motion act related to an assigned configuration it corresponds a power $W=\sum_{\nu}\mathbf{F}_{\nu}\cdot\dot{\mathbf{r}}_{\nu}$ of the forces; if the forces system is:
\begin{enumerate}
\item an active force system $\mathbf{F}_{a\nu}$ then force laws related to positions and velocities are imposed, obtaining the consequent virtual power of the active forces $W_{a}^{(v)}$
\item a virtual motion act with the constrain reaction system  $\mathbf{F}_{r\nu}$, then a virtual power of the reactive forces $W_{r}$ is considered. 
\end{enumerate}
It follows the definition:

\begin{definition}
- \emph{[\ref{morro}]}
A holonomic system is perfect or with perfect constrain if the virtual power of the reactive forces is zero for all virtual motion act.
\end{definition}

\begin{comment}
- \emph{[\ref{morro}]}
If a point is constrained to a movable surface, the smoothness of the constrain can be expressed not using the zero constrain reaction power, but the zero virtual power, because it corresponds to the velocity vectors tangent to the surface at the considered time. This issue can not be used for scleronomic constrains.
\end{comment}

\begin{comment}
- \emph{[\ref{morro}]}
The virtual power of the active forces is a linear form of the components $\delta q_{i}$ whose coefficients are defined as lagrangian forces or lagrangian components of the active forces:
\begin{equation} 
W_{a}^{(v)}=\sum_{\nu}\mathbf{F}_{a\nu}\cdot\delta\mathbf{r}_{\nu}=\sum_{\nu}\mathbf{F}_{a\nu}\cdot\sum_{i}\frac{\partial\mathbf{r}_{\nu}}{\partial q_{i}}\delta q_{i}=\sum_{i}\varphi_{i}\delta q_{i}
\end{equation}
form which
\begin{equation}
\varphi_{i}=\sum_{\nu}\mathbf{F}_{a\nu}\cdot\frac{\partial \mathbf{r}_{\nu}}{\partial q_{i}}
\end{equation}
If the active forsec are functions of the positions and of the velocities then the lagrangian forces are $\varphi =\varphi (\mathbf{q},\dot{\mathbf{q}},t)$.
\end{comment}

\begin{definition}
- \emph{[\ref{morro}]}
The dynamic state of a mechanical system is the time distribution of the positions, velocities and accelerations of the points of the system.
\end{definition}

A virtual power $W_{m}^{(v)}$ of the mass forces, called also inertial forces, is associated to each dynamic state.  The insertial forces are defined by the Newton Law $\mathbf{F}_{m\nu}=-m_{\nu}\mathbf{a}_{\nu}$; the virtual power of the inertial forces is a linear form of the components $\delta q_{i}$ of the virtual motion act, too:
\begin{equation}
W_{m}^{(v)}=\sum_{\nu}\mathbf{F}_{m\nu}\cdot\delta\mathbf{r}_{\nu}=-\sum_{\nu}m_{\nu}\mathbf{a}_{\nu}\cdot\sum_{i}\frac{\partial\mathbf{r}_{\nu}}{\partial q_{i}}\delta q_{i}=\sum_{i}\tau _{i}\delta q_{i}
\end{equation}
form which it follows
\begin{equation}
\tau _{i}=-\sum_{\nu}m_{\nu}\mathbf{a}_{\nu}\cdot\frac{\partial\mathbf{r}_{\nu}}{\partial q_{i}}
\end{equation}

\begin{statement} - \emph{\textbf{Lagrange-D'Alembert Principle} [\ref{morro}]} - 
In any dynamic state of a system with perfect constrains, for all virtual motion acts the sum of the virtual powers of the active forces and of the inertial forces equals zero:
\begin{equation}
W_{a}^{(v)}+W_{m}^{(v)}=0
\end{equation}
\end{statement}

\begin{definition}
- \emph{[\ref{morro}]}
A system $(Q,\mathcal{L})$ is said lagrangian if it is a differential variety $Q$ of dimension $n$, said configuration variety, with an associated real function $\mathcal{L}:TQ\times\mathbb{R}\rightarrow\mathbb{R}$. If the system is time independent the lagrangian is a function $\mathcal{L}:TQ\rightarrow\mathbb{R}$. The lagrangian dynamics is the set of curves expressed by the first order system of $2n$ differentiable equations:
\begin{equation}
\left\{
\begin{array}{ll}
\dot{q}_{i}=\dfrac{dq_{i}}{dt}
\\
\dfrac{d}{dt}\bigg(\dfrac{\partial\mathcal{L}}{\partial\dot{q}_{i}}\bigg)-\dfrac{\partial\mathcal{L}}{\partial q_{i}}
\end{array}
\right.
\label{eulero-lagrange}
\end{equation}
where the \emph{(\ref{eulero-lagrange})$_{2}$} equations are the Euler-Lagrange ones.
\end{definition}

\begin{comment}
- \emph{[\ref{morro}]}
For the holonomic systems the intrinsic properties of the Euler-Lagrange equations, i.e. the independence of the lagrangian coordinates choose, is the consequence of application of the Lagrange-D'Alembert principle to a lagrangian equation system.
\end{comment}

\begin{definition}
- \emph{[\ref{morro}]}
A functional is an application $\phi :\Omega\rightarrow\mathbb{R}$ such that for all $n-$tupla of functions corresponds a real number. A functional is differentiable in a point $q_{i}(t)\in\Omega$ if for all the chooses of the growth, said cariations, $\delta q_{i}(t)\in\Omega$ there exists the following relation:
\begin{equation}
\phi (q_{i}+\delta q_{i})=\phi (q_{i})+\delta\phi (q_{i},\delta q_{i})+\mathcal{R}
\end{equation}
where $\delta\phi$ is a linear functional of $\delta q_{i}$ and $\mathcal{R}$ is a functional of upper order in the same increases.
\end{definition}

\begin{definition}
- \emph{[\ref{morro}]}
A variation $\delta q_{i}(t)$ is saied end fixed if:
\begin{equation}
\delta q_{i}(t_{1})=\delta q_{i}(t_{2})=0
\label{estremifissi}
\end{equation}
\end{definition}

\begin{definition}
- \emph{[\ref{morro}]}
The action is defined as:
\begin{equation}
\mathcal{A}=\int_{t_{1}}^{t_{2}}\mathcal{L}\big(t,q_{i}(t),\dot{q}_{i}\big)dt
\end{equation}
\end{definition}

\begin{theorem}
- \emph{[\ref{landau},\ref{morro}]} - \emph{\textbf{Least action principle}}.\\
The function $q_{i}(t)$ for which $\delta\mathcal{A}=0$ for all \textbf{end fixed variations}, are only the solutions of the differential system \emph{(\ref{eulero-lagrange})}, where the lagrangian is defined up to a function of the coordinates and time.
\end{theorem}

A general approach to the mechanical systems can be developed using the least action principle, said also Hamilton principle, for which the mechanical system is described using a lagrangian function $\mathcal{L}(\mathbf{q};\dot{\mathbf{q}},t)$ form which the action can be obtained [\ref{landau}]:
\begin{equation}
\mathcal{A}=\int_{t_1}^{t_2}\mathcal{L}(\mathbf{q};\dot{\mathbf{q}},t)
\end{equation}
The Hamilton principle states that the motion of a system follows the path $\mathbf{q}(t)$ for which the action is minimum:
\begin{equation}
\delta\mathcal{A}=\delta\int_{t_1}^{t_2}\mathcal{L}(\mathbf{q};\dot{\mathbf{q}},t)=0
\end{equation}

The proof of this relation can be obtained starting from the hypothesis that the least value of the action is $\mathbf{q}(t)$ and a small variation $\delta\mathbf{q}$ around it is considered. Then for
\begin{equation}
\mathbf{q}(t) + \delta\mathbf{q}(t)
\label{posizione}
\end{equation}
the action $S$ growths [\ref{landau}], but for $t=t_1$ and $t=t_2$ the relation (\ref{posizione}) must have the fixed values $\mathbf{q}(t_1)=\mathbf{q}_1$ and $\mathbf{q}(t_2)=\mathbf{q}_2$, fundamental conditions for the Hamilton principle [\ref{landau}]:
\begin{equation}
\delta\mathbf{q}(t_1)=\delta\mathbf{q}(t_2)=0
\label{condizione1}
\end{equation}

Consequence of the least action principle is the Lagrange differential equations system:
\begin{equation}
\frac{d}{dt}\frac{\partial\mathcal{L}}{\partial \dot{q}_i}-\frac{\partial\mathcal{L}}{\partial q_i}=0 \qquad i\in[1,n]
\end{equation}

\section{Nonholonomic constrains}
A free point $P$, from any initial position $P_{0}$ at the initial time $t_{0}$, can move of an elementary displacement $dP=\mathbf{v}dt$; for a constrained point these displacements are confined because of the constrain [\ref{civita1}]. A holonomic system in a initial configuration at the time $t_0$, can have a transition to another configuration at the time $t_0+dt$ infinitely near to the initial one [\ref{civita1}].

\begin{definition} - \emph{[\ref{civita1}] }
A possible displacement at the time $t$, starting from a configuration $C$, is any infinitesimal displacement of a honolomic system which allows it to have a transition from the configuration $C$ at the time $t$ to a configuration $C'$ at the time $t+dt$:
\[
P_i=P_i(\mathbf{q};t)\mapsto P_i+dP_i=P_i(\mathbf{q}+d\mathbf{q};t+dt)
\]
from which the possible displacement are the $n$ equations:
\begin{equation}
dP_i=\sum_{k}\frac{\partial P_i}{\partial q_k}dq_k+\frac{\partial P_i}{\partial t}dt=\nabla_{\mathbf{q}} P_{i}\cdot d\mathbf{q}+\frac{\partial P_i}{\partial t}dt
\end{equation}
\end{definition}

If the virtual displacement are coupled the holonomic constrains equations (\ref{olonomi}) related to the displacements themselves, represented by the $l$ equations:
\begin{equation}
df_j=\sum_{k}\frac{\partial f_j}{\partial q_k}dq_k+\frac{\partial f_j}{\partial t}dt=\nabla_{\mathbf{q}}  f_{j}\cdot d\mathbf{q}+\frac{\partial f_j}{\partial t}dt=0
\end{equation}
only $n-l$ free lagrangian coordinates can be obtained. Dividing for $dt$ the nonholomic constrain relation can be obtained (\ref{anolonomi}):
\begin{equation}
\frac{df_j}{dt}=\sum_{k}\frac{\partial f_j}{\partial q_k}\dot{q}_k+\frac{\partial f_j}{\partial t}=\nabla_{\mathbf{q}}  f_{j}\cdot\dot{\mathbf{q}}+\frac{\partial f_j}{\partial t}=\sum_{k}a_{jk}\dot{q}_k+b_j=(\mathbf{a}\cdot\dot{\mathbf{q}}+b)_j=0
\label{anol}
\end{equation}
which is a restrain in the motion. So the displacement are limited and the virtual displacements must be introduced:

\begin{definition} - \emph{[\ref{civita1}] }
A virtual displacement is any hypothetic displacement which allows the system to have a transition from a configuration $C$ to another infinitesimal near one $C'$ allowed by the constrains at the same time.
\end{definition}

Consequently, for the nonholonomic constrains $dt=0$ and the relation (\ref{anol}) becomes:
\begin{equation}
\nabla_{\mathbf{q}}  f_{j}\cdot d\mathbf{q}+\frac{\partial f_j}{\partial t}dt=0\Rightarrow\mathbf{a}\cdot\delta{\mathbf{q}}=0
\label{anol1}
\end{equation}

\begin{definition} - \emph{[\ref{civita2}] }
The live force is the value of the kinetic energy.
\end{definition}

\begin{theorem} - \emph{[\ref{civita2}] }
\emph{\textbf{Theorem of live forces or of K\"onig}} - The live force of any system in motion is the sum of the live force of the centre of mass and the one of the motion in relation to the centre of mass.
\end{theorem}

Volterra pointed out that the lagrangian is an explicital function of $\dot{\mathbf{q}}$ [\ref{civita2}].

\section{The answer proposed}
The variational methods are fundamental in the development of modern analytical mechanics [\ref{flannery}], but Flannery pointed out [\ref{blochfirst},\ref{blochlast}]:
\begin{quotation}
the least action principle can be applied only to holonomic and linear nonholonomic constraints, while it is not useful to obtain the correct equations of motion for general nonholonomic constraints.
\item 
\end{quotation}

The answer here proposed is the following. The basis of the least action principle is the evaluation of the variations based on the hypothesis of the fix ends (\ref{estremifissi}) [\ref{elsgolts}]. For the nonholonomic constrains (\ref{anol}) and (\ref{anol1}) can be easily proved that at least one of the virtual displacements can be written as a linear combination of the others; i.e., 
\begin{equation}
\delta q_i(t)=a_{i}^{-1}\sum_{j}a_{ij}\delta q_j
\end{equation}
Consequently, the relations (\ref{anol}) e (\ref{anol1}), based of the least action principle, are not satisfied. For the nonholonomic constrains the fundamental conditions of use of the least action principle are not true.

\bibliographystyle{unsrt}

\end{document}